\documentclass[10pt, journal]{IEEEtran}

\usepackage[utf8]{inputenc}
\usepackage[T1]{fontenc}
\usepackage{amsmath}
\usepackage{graphicx}
\usepackage{color}
\usepackage{cite}

\begin{document}
\title{Electromagnetic design of a superconducting electric machine with bulk HTS material}

\author{Roman~Bause, Mark~D.~Ainslie,~\IEEEmembership{Senior~Member,~IEEE,} Matthias~Corduan, Martin~Boll, Mykhaylo~Filipenko, and~Mathias~Noe%
\thanks{R.\ Bause was with Siemens AG, Corporate Technology, 82024 Taufkirchen, Germany, during the preparation of this work. He is now with Max Planck Institute of Quantum Optics, 85741 Garching, Germany (e-mail: roman.bause@mpq.mpg.de).}%
\thanks{M.\ D.\ Ainslie is with the Department of Engineering, University of Cambridge, Cambridge CB2 1PZ, United Kingdom.}%
\thanks{M.\ Boll, M.\ Corduan and M.\ Filipenko are with Siemens AG, Corporate Technology.}%
\thanks{M.\ Noe is with the Institute of Technical Physics, Karlsruhe Institute of Technology, 76344 Eggenstein-Leopoldshafen, Germany.}}%

\maketitle

\begin{abstract}
The use of high-temperature superconductors in electric machines offers potentially large gains in performance compared to conventional conductors, but also comes with unique challenges. Here, the electromagnetic properties of superconducting electric machines with bulk HTS trapped-field magnets in the rotor and conventional copper coils in the stator are investigated. To this end, an analytical model of the electromagnetic field in radial air-gap synchronous electric machines is developed and validated, taking into account the specific difficulties that occur in the treatment of machines with bulk HTS. Using this model, the influence of pole pair number, stator winding thickness, rotor surface coverage, and air gap width on the machine's Esson coefficient is calculated. In contrast to numerical simulations, the method presented here can provide results within minutes, making it particularly useful for work in early development and systems engineering, where large parameter spaces must be investigated quickly.
\end{abstract}

\section{Introduction}
Commercial interest in high-temperature superconducting (HTS) electric machines has recently intensified. This is because HTS materials enable a considerable increase in air gap field and current loading, leading to higher torque and power output than would be possible in a conventionally-conducting machine of the same size and weight. HTS machines could be used for applications where conventional electric machines can not meet the performance requirements such as hybrid-electric passenger aircraft~\cite{luongo2009, berg2015, haran2017, sanchez2017, durrell2018}. However, compared to conventional electric machines, which have been studied extensively, HTS machines behave differently in many ways. Having an analytical model of the electromagnetic characteristics of the machine available can simplify and speed up the design, because with analytical calculations, results can be obtained very quickly and larger parameters spaces can be investigated than would be possible with more time-consuming finite-element calculations. This can be particularly useful in the context of system optimization, e.g. to determine the influence of machine properties on other parts of a drive-train such as power sources or power distribution~\cite{brown2011}.

In this paper, we develop an analytical model for the electromagnetic field in the HTS equivalent of a radial-gap permanent magnet synchronous machine in which bulk HTS material is used to produce the magnetic field instead of conventional permanent magnets in the rotor. While conventional permanent magnets are limited to fields of 1.2 T in practical applications~\cite{gieras2008}, the field trapped in bulk HTS material can be significantly larger. Furthermore, experiments have shown that using bulk HTS magnets instead of conventional permanent magnets in synchronous machines is possible despite the challenges of cooling and magnetization~\cite{matsuzaki2005, jiang2008, deng2011, zhou2012, zhang2016}.

\section{Bulk HTS Materials}
\label{bulk-hts-material}
Progress continues to be made on improving fabrication techniques for bulk HTS materials, and fields in excess of 17~T have been achieved in bulk HTS magnets with the field cooling magnetization (FCM) method~\cite{durrell2014}. One significant challenge to the usage of bulks in practical applications is achieving a simple, reliable and compact magnetization process, which essentially involves the application and removal of a large magnetic field that induces a circulating supercurrent in the bulk. The bulk may then act as a super-strength, quasi-permanent magnet. Pulsed field magnetization (PFM) shows the greatest promise as a practical magnetization method, where the large external magnetization field is applied via a pulse with a duration on the order of milliseconds, and the stator coils of a machine can be exploited for this purpose~\cite{zhang2016}. However, the trapped field generated via PFM is generally much smaller than the true capability of the material as indicated by FCM due to the large temperature rise associated with the rapid dynamic movement of magnetic flux in the interior of the material~\cite{ainslie2015}. To date, the record trapped field by PFM is 5.2~T at 29~K~\cite{fujishiro2006}. Recently, magnetic fields >~3~T at temperatures above 50~K have been trapped using PFM and a portable Stirling cryocooler setup~\cite{zhou2017, yokoyama2018}. It has also been shown that trapped-field magnets can be improved in regards to stability against demagnetization by external AC fields~\cite{srpcic2018} and against mechanical forces~\cite{huang2018}. In previously tested machines, the bulk HTS magnets were only magnetized to 1.3 T or less~\cite{sano2008, watasaki2013, huang2014, huang2016}. However, it is reasonable to assume that with the application of more optimized PFM methods as well as a machine geometry specifically designed to allow high magnetization, this number can be increased to reach values similar to the state of the art for PFM in a laboratory environment.

\begin{figure}[!t]
\centering
\includegraphics[width=3.5in]{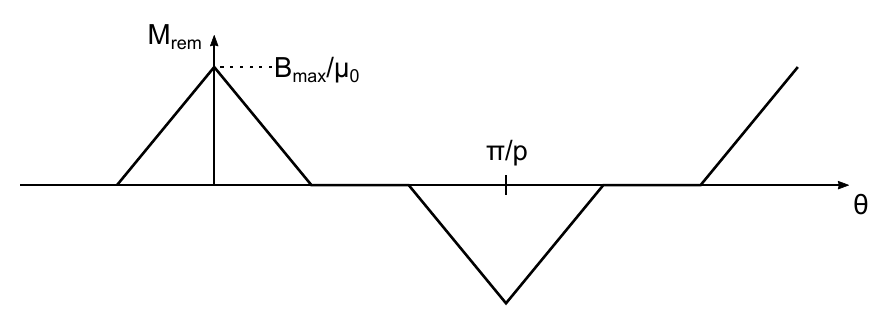}
\caption{Sketch of the remanent magnetization in the machine, as given by (\ref{remanence-in-magnets}).}
\label{triangle}
\end{figure}

The circulating supercurrent in a disc-shaped bulk superconductor gives rise to a conical trapped magnetic field profile above the top surface of the bulk in 3D, which can be approximated to a triangular profile in 2D~\cite{bean1964, shen2015}. This profile is shown in Fig.~\ref{triangle}. This is in contrast to models for conventional permanent magnet machines, where a rectangular field distribution is usually used. 

For the purposes of our model, we treat the trapped magnetic field as a remanent magnetization, analogous to a permanent magnet. While this does not reflect the microscopic mechanism by which the field is trapped, it accurately reproduces the real magnetic field distribution. The Fourier representation of the triangular profile is given by

\begin{align}
\label{remanence-in-magnets}
M_{\mathrm{rem},r} = \sum_{n \: \mathrm{odd}} 2 \alpha_M M_{\text{max}, r} \left(\operatorname{sinc}\left(\frac{n\alpha_M}{4}\right)\right)^2 \cos(np\theta),
\end{align} 

where $M_{\text{rem},r}$ denotes the component in the radial direction of the remanent magnetization, $\alpha_M$ is the duty cycle, i.e.,\ the fraction of the rotor surface which is covered by magnets, and $M_{\text{max}, r}$ is the maximum remanent magnetization in the radial direction. The $\text{sinc}$ function is defined as $\operatorname{sinc}(x) = \sin(x) / x$.

\section{Analytical Model}
Although a number of analytical and numerical models have been developed to describe various aspects of bulk superconductor magnetization~\cite{ainslie2015}, there has been little development on bulk superconductor-based electric machine modelling~\cite{haran2017}. 

The simplest way of analytically modelling the field in electric machines is to assume a very small air gap surrounded by iron, where the magnetic field only has a component in the radial direction and is constant in this direction. These assumptions do not make sense in a bulk HTS machine because the rotor must be surrounded by thermal insulation, leading to a large air gap where significant change of the magnetic field in the radial direction can occur. We therefore apply a two-dimensional approach, in which the magnetic field depends both on the radial and angular position in the air gap.

\begin{figure}[!t]
\centering
\includegraphics[width=3.49in]{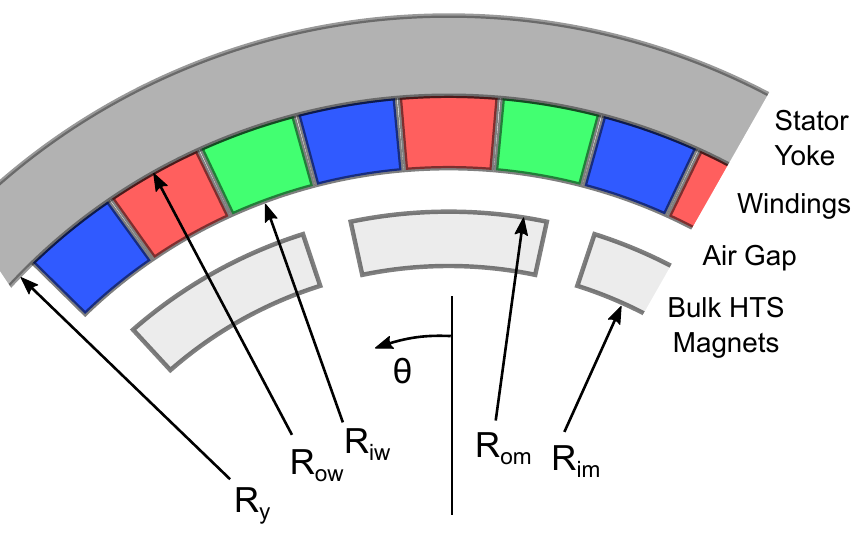}
\caption{Sketch of the machine dimensions as used for the analytical model. The three phases of the stator winding are shown in red, green, and blue.}
\label{geometry}
\end{figure}

In the following, general equations for the magnetic field in a cylindrically symmetric machine in polar coordinates $(r, \theta)$ are derived. To do so, the machine geometry is divided into zones where the relative permeability $\mu_r$ is assumed to be constant. Each zone is ring-shaped, i.e.\ $R_i \leq r < R_o$, $0 \leq \theta < 2 \pi$ for some inner and outer radii $R_i$ and  $R_o$, respectively. The geometry used for the model is shown in Fig.~\ref{geometry}. In each zone, the relationship between the fields $\boldsymbol{B}$ and $\boldsymbol{H}$ and the remanent magnetization $\boldsymbol{M}_\mathrm{rem}$ is given by

\begin{equation}
 \boldsymbol{B} = \mu_0 \left( \mu_r \boldsymbol{H} + \boldsymbol{M}_\mathrm{rem} \right).
\end{equation}

\subsection{Open-Circuit Field}
\label{open-circuit-field}
As a first step, the magnetic field of the bulk HTS magnets in the rotor is computed, neglecting the influence of the stator windings. The zones are defined as follows: Zone 1 is the inner part of the rotor, which is assumed to consist of a non-magnetic material ($\mu_r=1$). Zone 2 contains the bulk HTS magnets with $M_{\mathrm{rem},r}$ given by~(\ref{remanence-in-magnets}). Zone 3 contains both the air gap and the stator windings, since the permeability $\mu_r$ of most winding materials is close enough to unity that the effect of it can be safely neglected. Finally, zone 4 is the stator yoke made of a highly permeable material, so we assume $\mu_r \to \infty$, which is justified if the distance between the rotor magnets and the yoke is large enough for the field to decay below the yoke's saturation field. An overview of the zone layout is given in Table~\ref{zone-layout}. Because we treat the bulk HTS magnets like permanent magnets, it is possible to use a scalar magnetic potential which is defined by $\boldsymbol{H} = - \nabla \phi$, to calculate the fields. Using an approach analogous to~\cite{zhu1993a}, the Poisson equation 

\begin{align}
\label{poisson-equation}
\nabla^2 \phi = \frac{\partial^2  \phi}{\partial r^2} + \frac{1}{r} \frac{\partial \phi}{\partial r} + \frac{1}{r^2} \frac{\partial \phi^2}{\partial^2 \theta} = \frac{M_{\mathrm{rem}, r}}{\mu_r r}
\end{align}

can be solved to find $\phi(r,\theta)$ for each zone. The solution is given in Appendix~\ref{potential-solution}. The resulting field components in zone $m$ are

\begin{align}
B&_{r,m}(r, \theta) = -\mu_0 \mu_{r,m} \sum_{n \: \mathrm{odd}} np M_{n} \Big(   A_{n,m} r^{np -1}\nonumber \\ 
& - C_{n,m} r^{- np - 1} + \frac{\delta_{m,2} \, np}{\mu_{r,m} (1-\left(np\right)^2)}\Bigg)\cos(np\theta), \\ 
B&_{\theta,m}(r, \theta) = \mu_0 \mu_{r,m} \sum_{n \: \mathrm{odd}} np M_{n} \Big(A_{n,m} r^{np -1} \nonumber\\
& + C_{n,m} r^{- np - 1} + \frac{\delta_{m,2}}{\mu_{r,m} (1-\left(np\right)^2)}\Bigg) \sin(np\theta).
\end{align}

Here, $n$ denotes the harmonic order, $p$ is the pole pair number of the machine, $M_n$ are the Fourier coefficients from (\ref{remanence-in-magnets}) and $\delta_{n,m}$ is the Kronecker symbol, defined as $\delta_{n,m} = 1 $ for $n=m$ and 0 otherwise. The coefficients $A_{n,m}$ and $C_{n,m}$ depend on the machine geometry and are found by imposing the condition that $B_r$ and $H_\theta$ must be continuous across all zone boundaries. These coefficients are listed in Appendix \ref{coefficients-open-circuit}. 

A method to deal with the singular case $np=1$ can be found in~\cite{zhu1993a}, however, in this study,  we limit ourselves to the more common case of $p>2$.

\begin{table}[!t]
\renewcommand{\arraystretch}{1.3}
\caption{Definition of zones}
\label{zone-layout}
\centering
\begin{tabular}{c c c}
\hline
\bfseries Zone & \bfseries Open-circuit & \bfseries Armature reaction\\
\hline
1 & $r < R_{im}$ & $ r < R_{iw}$ \\
2 & $R_{im} \leq r < R_{om}$ & $R_{iw} \leq r < R_{ow}$ \\
3 & $R_{om} \leq r < R_{y}$ & $R_{ow} \leq r < R_{y}$ \\
4 & $R_y \leq r$ & $R_y \leq r$ \\
\hline
\end{tabular}
\end{table}

\subsection{Armature Reaction Field}

In the second step, the field created by the stator windings is computed. For this computation, a different zone distribution is required: Zone 1 encompasses the entire space inside the stator (i.e.,\ the rotor and air gap), which is assumed to have $\mu_r = 1$. Zone 2 contains the windings. Zone 3 is the space between the stator windings and yoke, which is assumed to have zero width in all subsequent calculations and is therefore not shown in Fig.~\ref{geometry}. Zone 4 is the stator yoke with $\mu_r \to \infty$. Instead of the scalar potential $\phi$, the vector potential $\boldsymbol{\varphi}$ defined by $\boldsymbol{H} = \nabla \times \boldsymbol{\varphi}$, must be used for this calculation, because currents have to be taken into account. For a two-dimensional model of the magnetic field, only the component along the machine axis, $\varphi_z$, is relevant. Therefore, when replacing $\phi$ with $\varphi_z$, a modified version of (\ref{poisson-equation}) is still valid:

\begin{equation}
\nabla^2\varphi_z = - J,
\end{equation}

where $J$ is the current density. An analogous approach to the one shown in the previous chapter can be used to find the solution of this differential equation~\cite{woodson1966, liu2018}. The general solution for $\varphi_z$ is shown in Appendix \ref{potential-solution}. The magnetic field can be computed via $B_r(r, \theta) = \mu_0 \mu_r r^{-1} \cdot \partial \varphi_z / \partial \theta$ and $B_\theta(r, \theta) = - \mu_0 \mu_r \partial \varphi_z / \partial r$. 

The winding configuration we consider is a distributed winding with $N_p$ phases as seen in Fig.~\ref{geometry} for the case $N_p=3$. Within the cross-section of the conductors of each phase, we always assume constant $J$. Therefore, the Fourier components of the spatial current distribution along the tangential machine direction are

\begin{equation}
\label{j-fourier}
J(\theta) = \sum_{n \: \mathrm{odd}} 2 \alpha J_{\mathrm{max}} \mathrm{sinc}\left( \frac{n \alpha}{2} \right) \cos(np\theta).
\end{equation}

The components of the armature reaction field in zone $m$ for phase $k$ are

\begin{align}
B&_{r,m}(r, \theta, t) = -\mu_0 \mu_{r,m} J_k(t) \sum_{n \: \mathrm{odd}} F_n np \Big( A'_{n,m} r^{np -1} \nonumber \\ 
& + C'_{n,m} r^{- np - 1} - \frac{\delta_{m,2}\,r}{4-\left(np\right)^2}\Bigg)\sin\left(n\left(p\theta-\frac{2\pi k}{N_p}\right)\right), \\
B&_{\theta,m}(r, \theta, t) = -\mu_0 \mu_{r,m} J_k(t) \sum_{n \: \mathrm{odd}} F_n np \Big(A'_{n,m} r^{np -1} \nonumber\\
& - C'_{n,m} r^{- np - 1} - \frac{\delta_{m,2}\, 2r}{4-\left(np\right)^2}\Bigg) \cos\left(n\left(p\theta-\frac{2\pi k}{N_p}\right)\right).
\end{align}

\begin{figure}[!t]
\centering
\includegraphics[width=3.49in]{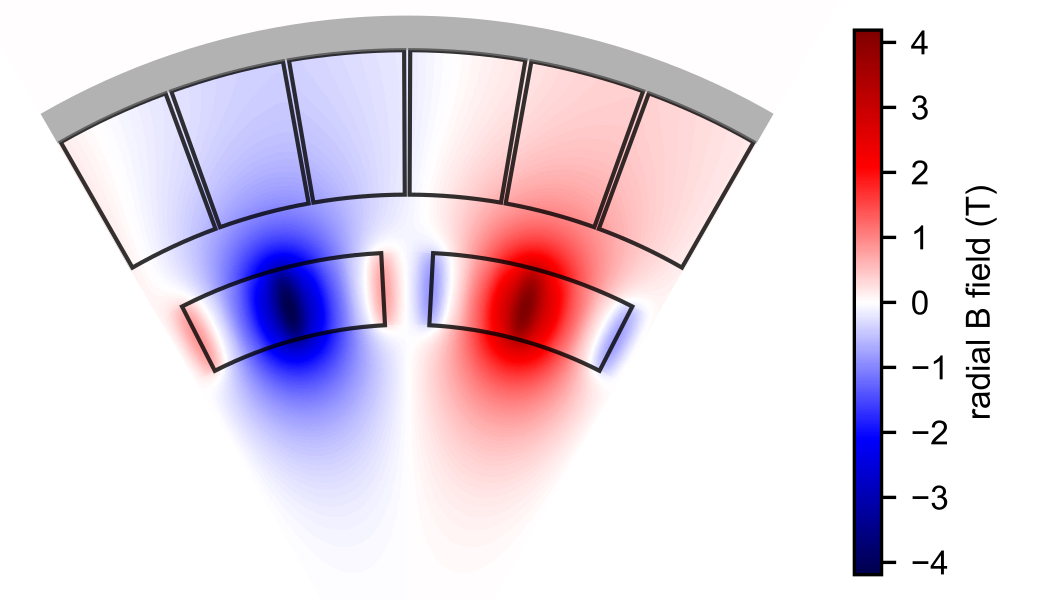}
\caption{Radial component of the total $B$ field, i.e.,\ the sum of the open-circuit and armature reaction fields, as computed with our model for the base-line machine described in Table~\ref{base-line-design}.}
\label{total_field}
\end{figure}

Here, $F_n$ is introduced as a shortcut for the Fourier components given in (\ref{j-fourier}) and $J_k(t)$ is the current density in phase $k$ at time $t$. The coefficients $A'_{n,m}$ and $C'_{n,m}$ are listed in Appendix \ref{coefficients-armature-reaction}. An example of the magnetic field this model yields can be seen in Fig.~\ref{total_field}.

\subsection{Computation of Torque}
The torque produced by a machine can be computed by using the equation for the Lorentz force,

\begin{equation}
F_L = \boldsymbol{J} \times \boldsymbol{B}.
\end{equation}

From this, with the assumption that the current density in the stator is purely in $z$-direction, it follows that the torque is given by

\begin{equation}
T(t) = l_{\mathrm{eff}} \int_0^{2\pi} d\theta \int_{R_{iw}}^{R_{ow}} dr \, r^2 J(r,\theta ,t) B_r(r,\theta ,t).
\end{equation}

Here, $l_{\mathrm{eff}}$ is the effective machine length. By writing the total $J(t)$ for all phases $k$ in terms of Fourier components in both the spatial and time domains,

\begin{align}
J(r, \theta, t) = \sum_{n,l,k} F_n & J_l \cos \left(2 \pi l \left(pft - \frac{k}{N_p}\right)\right) \nonumber \\ &\times \cos \left(n \left(p\theta - \frac{2 \pi k}{N_p}\right)\right),
\end{align}

and making use of the orthogonality of the cosine functions, the mean torque $\overline{T}$ can be expressed as

\begin{align}
\label{torque}
\overline{T} &= f_{\mathrm{mech}} \int_0^{1/f_{\mathrm{mech}}} T \,  dt  = -\frac{\pi}{2} \mu_0 N_p l_{\mathrm{eff}} \sum_{n \: \mathrm{odd}} F_n M_n J_n np \nonumber \\ & \hspace{2cm} \times \int_{R_{iw}}^{R_{ow}} dr \, A_{n,3} r^{np+1} - C_{n,3} r^{-np+1}.
\end{align}

From this, the Esson coefficient $C$ for a given rotor radius $R_{om}$ and machine length $l$ can by computed: 

\begin{equation}
\label{esson}
C = \frac{\pi \overline{T}}{2 R_{om}^2 l} 
\end{equation}

We use this number to estimate the effectiveness of the electromagnetic part of a machine design in terms of efficient use of available space. This offers a convenient way to compare the electromagnetic characteristics of different machines. Note that $C$ is usually given in units of  W$\,$min/m$^3$.
Typical values for air-cooled machines at nominal powers on the order of 100 kW are 4 to 5 kW$\,$min/m$^3$~\cite{binder2012}.
\subsection{Implementation}
The model was implemented in the \textit{python} programming language, making use of the \textit{sympy 1.1.1} and \textit{numpy 1.12.x} libraries for analytical and numerical computations~\cite{jones2001}. Harmonic orders $n \leq 19$ were computed for all model runs. We were able to solve the analytical equations and evaluate them numerically over $10^6$ data points within one minute with a 3.2 GHz CPU.

There are two cases where two zones with the same permeability are adjacent. In these cases, a model with fewer zones could be used, or else the permeability must be made slightly different in one of the zones to avoid a singularity of the model. To preserve the generality of the model, we used the latter approach and assumed $\mu_r = 1+\epsilon$ in the case of the bulk HTS magnets (zone 2 in the model of the open-circuit field), where $\epsilon$ is much smaller than the precision of the model. In the case of the copper windings (zone 2 for the armature reaction field), we took into account the fact that copper is weakly diamagnetic and assumed a value slightly less than 1.  

\section{Electromagnetic Machine Design}
\subsection{Geometrical Parameters}
To assess the performance and usefulness of the model, we performed calculations for an example machine. We computed the dependence of the Esson coefficient as given in (\ref{esson}) on certain parameters, while other parameters were set to values we consider realistic. The bulk HTS magnets were assumed to be cylindrical discs of 1.5 cm thickness and 4 cm diameter with a trapped field of 3 T at the surface center, similar to what was demonstrated in~\cite{zhou2017}. This assumption is motivated by the conditions imposed by manufacturing and magnetising the bulks: As long the material can be magnetized perfectly, larger bulk HTS disks are always favourable, since they can trap higher magnetic fields in their center. However, larger disks are more likely to contain material defects and require stronger external fields to fully magnetize. Since the magnet size is constant, in simulations with different values for $p$ or $\alpha_M$, the rotor radius is changed accordingly.

For the stator, we assumed a standard three-phase distributed stator winding made of copper wire to keep the design simple. The duty cycle $\alpha_W$ for each stator winding was chosen to be $0.95/N_p$, where $N_p$ is the phase number, in order to allow sufficient space between windings for insulation. For each phase, we assumed a sinusoidal time-dependence of the current. Further, we assumed a peak current density of 9~A/mm$^2$ in the stator cross-section, which can be achieved by interspersing water cooling tubes with copper wires. In this configuration, a peak current of 25~A/mm$^2$ flows through the copper wires, which make up 35\% of the cross section of each winding. This configuration is similar to the state of the art in commercial non-superconducting, high power-density electric machines~\cite{galea2014}. In order to partially correct the error caused by using a two-dimensional model, the machine length $l$ was multiplied by a correction factor to obtain an effective length, $l_\mathrm{eff}$. The correction factor was found by comparing the total magnetic flux from an infinitely long bar of bulk HTS material with the flux from an infinitely long line of cylindrical bulks. The factor was found to be $\pi/6 \approx 0.5$. This correction factor does not take into account flux leakage in the axial direction, so that we can expect the model to overestimate the torque, with the error becoming larger for small machine lengths.

With these parameters set, we made preliminary decisions about the remaining geometric degrees of freedom according to experience with non-superconducting machines and some extrapolation, arriving at a base-line design, which served as a starting point for parameter scans. The properties of the base-line machine are summarized in Table~\ref{base-line-design}.

\begin{table}[tb]
\renewcommand{\arraystretch}{1.3}
\caption{Parameters and properties of the base-line machine}
\label{base-line-design}
\centering
\begin{tabular}{l l}
\hline 
\multicolumn{2}{c}{Model input}\\
Tangential magnet size & 4 cm \\
Magnet thickness & 1.5 cm \\
Peak air gap field $B_\mathrm{max}$ & 3 T \\
Rotation speed $f_{\mathrm{mech}}$ & 1500 rpm \\
Phase number $N_p$ & 3 \\
Machine length $l$ & 20 cm \\
Air gap width $d_A$ & 1.2 cm \\
Rotor radius $R_{om}$ & 10.2 cm \\
Stator winding current density $J$ & 9 A/mm${^2}$ \\
Rotor surface coverage $\alpha_M$ & 80 \% \\
Pole pair number $p$ & 6 \\
Stator winding thickness $d_W$ & 3.0 cm \\
\hline
\multicolumn{2}{c}{Model output}\\
Effective length $l_\mathrm{eff}$ & 10.5 cm \\
Mean torque $\overline{T}$ & 647 kN \\
Esson coefficient $C$ & 7.99 kW min/m$^3$ \\
Power $P$ & 102 kW \\
Estimated active part mass & 39 kg \\
\hline
\end{tabular}
\end{table}

\begin{figure}[tb]
	\centering
	\includegraphics[width=3.49in]{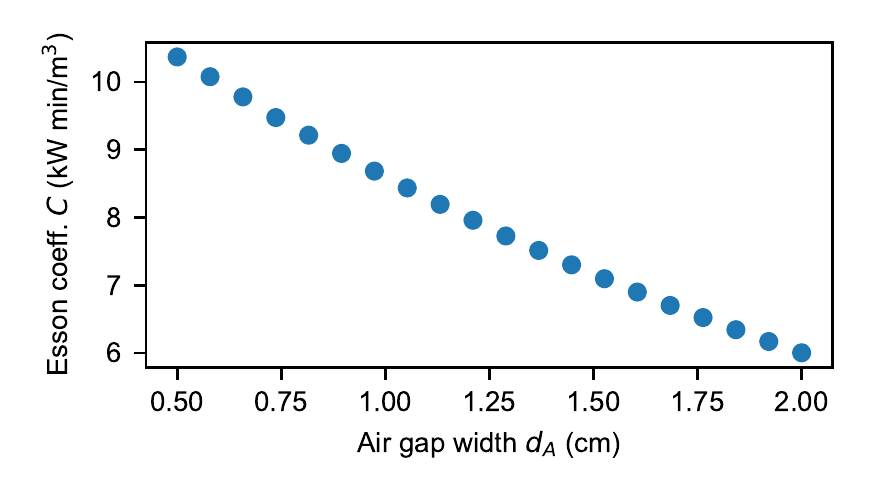}
	\caption{Esson coefficient $C$ versus air gap width $d_A$.}
	\label{c_vs_airgapwidth}
\end{figure}

Even though a smaller air gap is theoretically always beneficial, the air gap size
is limited by the requirement to accommodate cryogenic insulation around the rotor.
We expect that the smallest practical electromagnetic air gap width, with the rotor
operated at 50 K and the stator at room temperature, is 1.2 cm. A scan of the air gap width was performed to assess the importance of this parameter, and the results are shown in Fig.~\ref{c_vs_airgapwidth}. The large slope observed in this scan suggests that it may be worth developing advanced mechanical construction and isolation techniques to minimize air gap thickness, even though this is likely to be highly difficult.

\begin{figure}[tb]
	\centering
	\includegraphics[width=3.49in]{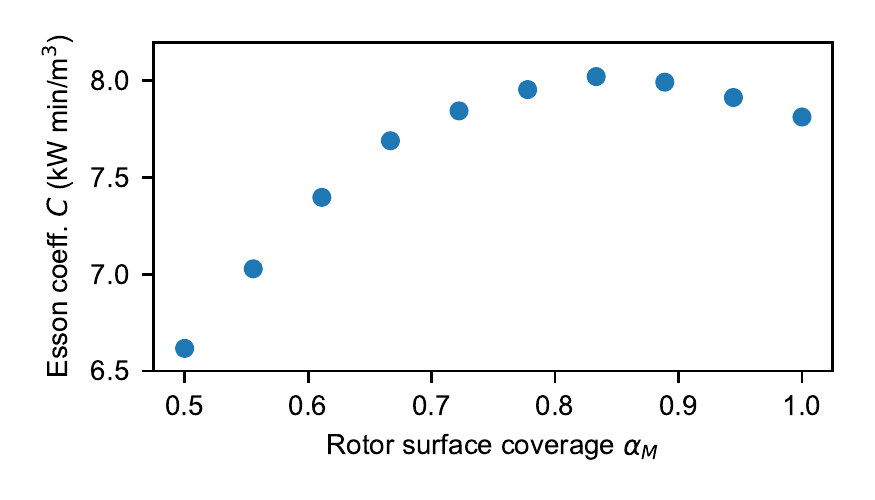}
	\caption{Esson coeffcient $C$ versus rotor surface coverage $\alpha_M$.}
	\label{c_vs_dutycycle}
\end{figure}

We also performed a scan of the impact of the rotor surface coverage $\alpha_M$ on the Esson number, as shown in Fig.~\ref{c_vs_dutycycle}. According to these data, increasing $\alpha_M$ beyond 85\% will actually decrease the torque. This is caused by the fact that, even though more total flux is produced on the rotor surface for higher $\alpha_M$, the leakage also increases and the additional flux never reaches the stator.

\begin{figure}[tb]
	\centering
	\includegraphics[width=3.49in]{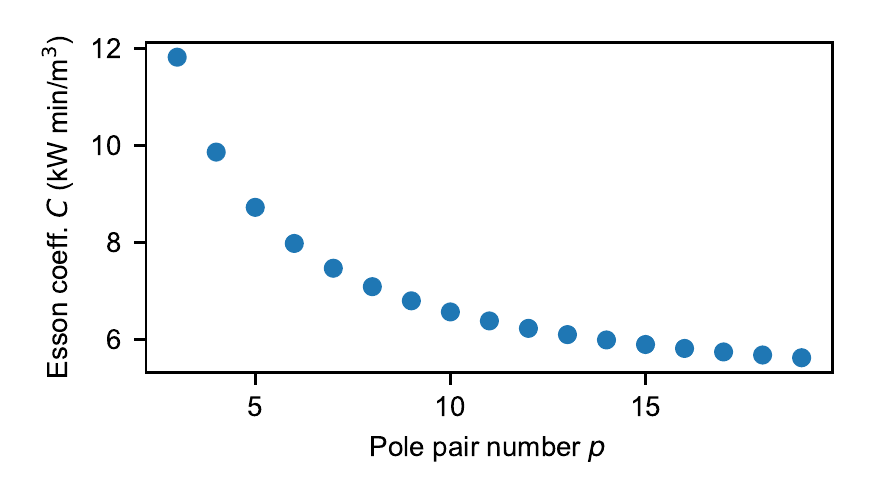}
	\caption{Esson coeffcient $C$ versus pole pair number $p$.}
	\label{c_vs_p}
\end{figure}

\begin{figure}[tb]
	\centering
	\includegraphics[width=3.49in]{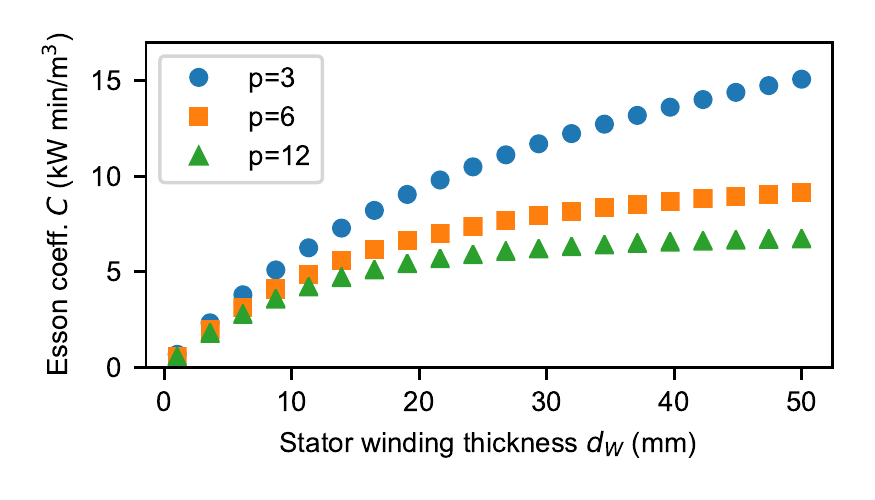}
	\caption{Esson coeffcient $C$ versus stator winding thickness $d_W$ for three different values of the pole pair number.}
	\label{c_vs_coilthickness}
\end{figure}

The two most interesting parameters to scan are the pole pair number $p$ and the thickness of stator windings $d_W$. This is because both of these strongly influence the machine's performance, while also containing trade-offs in the electromagnetic design. For example, one will always make the air gap width as small as is possible with given mechanical and thermal constraints, but $p$ and $d_W$ can behave non-linearly and will likely exhibit an optimum, even when only considering electromagnetic machine properties compared to the machine's volume or mass. As long as the stator windings are very thin, we can expect the generated torque to grow linearly with $d_W$, but this must saturate once the field inside the stator coils has decayed due to leakage. At this point, adding more copper would only contribute useless weight. We performed a scan of $p$ for the base-line design and and additional scan of $d_W$ at different values of $p$, to show the connection between these two variables, as shown in Figs. \ref{c_vs_p} and \ref{c_vs_coilthickness}. In the $p$-$C$ diagram, it can be seen that the Esson number first decays quickly and then approaches a constant value. This is because at low $p$, and correspondingly low machine radii, the stator windings are wedge-shaped, allowing for higher total currents and stator fields. At large $p$, the stator winding shape asymptotically approaches a rectangle.

\subsection{Model Validation}
\begin{figure}[tb]
	\centering
	\includegraphics[width=3.49in]{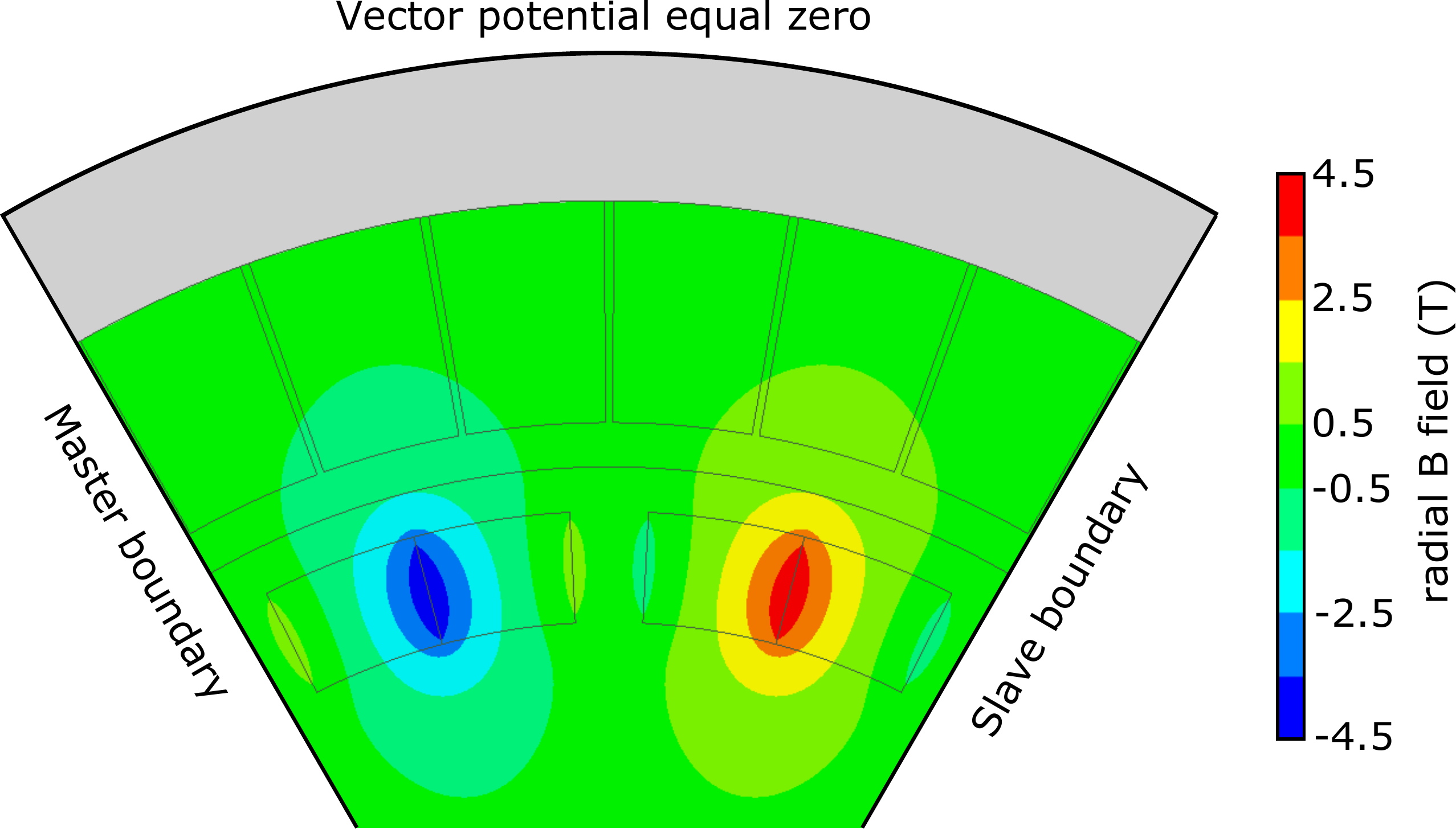}
	\caption{Radial component of the B field generated only by HTS magnets and computed with the parameters from the base-line machine in Table \ref{base-line-design} with FEM.}
	\label{Maxwell_OpencircuitField_Br}
\end{figure}

To validate our analytical model with a finite element method (FEM), we built a static two-dimensional solution type model in Ansys Maxwell 19.0.0, shown in Fig.~\ref{Maxwell_OpencircuitField_Br}. As input parameters, the model used the same dimensions and properties as for the particular case of the base-line machine shown in Table \ref{base-line-design}. Master and slave boundary conditions were set at the cut edges of the model, making use of symmetry, such that the simulation of only one twelfth of the machine is required. At the edge of the ambient air, the vector potential was set to zero. Power was supplied to the three-phase stator coils through solid current excitations with $5775 \,$A, which corresponds to 9$\,\mathrm{A/mm^2}$. 

The assumed typical triangular shaped field profile in the analytical model was mimicked in the FEM model by a single-turn coil of the same dimensions as the bulk HTS magnets. The supply and return conductor windings of this coil are adjoined and their current density was set to $346 \,\mathrm{A/mm^2}$ such that the resulting peak air gap field on the outer coil surface $B_\mathrm{max}$ reached the same value of $3 \,$T as for the HTS magnets.

\begin{figure}[tb]
	\centering
	\includegraphics[width=3.49in]{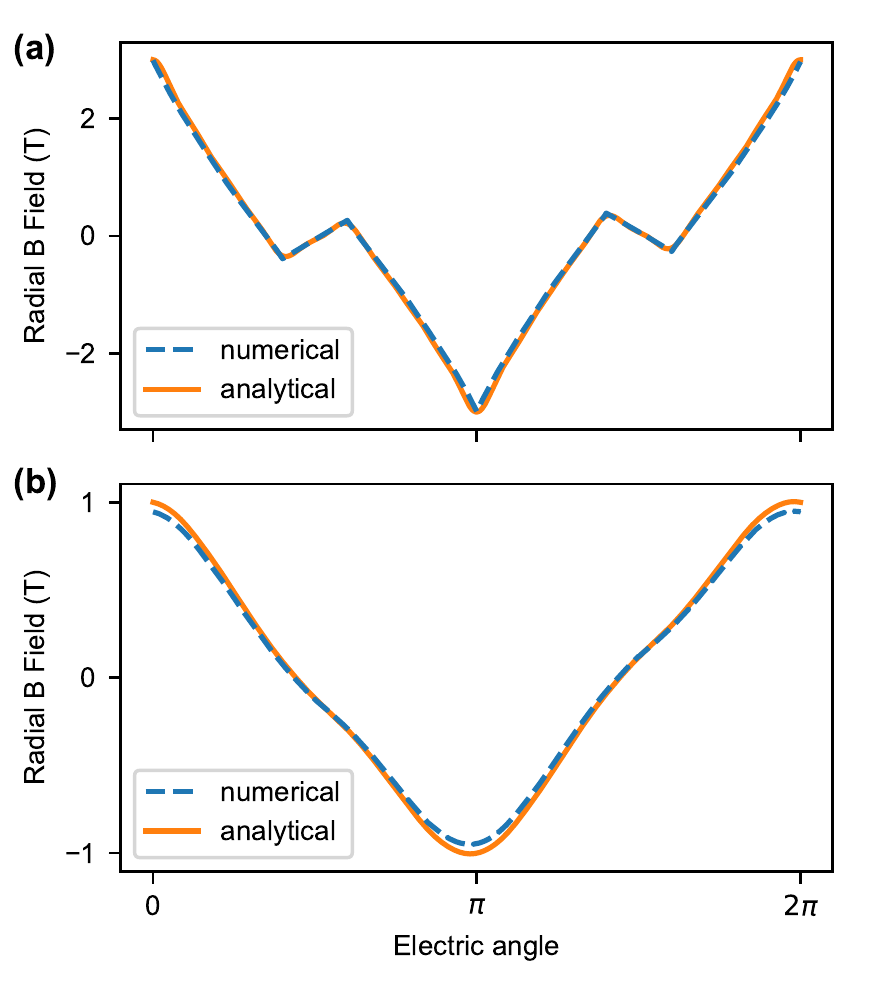}
	\caption{Radial field $B_r$ in the base-line machine as described in Table~\ref{base-line-design} in comparison of the analytical model and the numerical simulation. (a): At the outer radius of the magnets $R_{om}$  (b): At the inner radius of the stator windings $R_{iw}$.}
	\label{airgapfields}
\end{figure}

As seen in Fig. \ref{airgapfields}, the radial component of the total field matches the analytical results, with the small deviations seen close the field maxima at the stator surface being explained by the non-linear behaviour of the yoke. The machine achieved an Esson coefficient of 7.79$\, \mathrm{kW \, min/m^3}$ at a load angle of $\pi/$2 and is thus 2.5\% lower than in the analytical calculation. This level of agreement between the analytical and numerical models is to be expected due to the different approximations that were made.

\section{Acknowledgments}
M.\ A.\ would like to acknowledge financial support from an Engineering and Physical Sciences Research Council (EPSRC) Early Career Fellowship EP/P020313/1. All data are provided in full in the results section of this paper. We thank Marc Lessmann, Yingzhen Liu and John Durrell for helpful discussions.

\bibliographystyle{IEEEtran}
\bibliography{IEEEabrv,BulkSuperconductivity}

\newpage 
\appendices
\onecolumn

\section{Solutions of the Poisson Equation}
\label{potential-solution}
\begin{align}
&\phi_m(r,\theta) = \sum_{n \text{ odd}} M_n \left( A_{n,m} r^{np} + C_{n,m} r^{-np} + \frac{\delta_{m,2}\, r}{\mu_{r,m}\left(1-\left(np\right)^2\right)}\right) \cos ( np\theta), \\
&\varphi_{z,m}(r, \theta) = \sum_ {n \text{ odd}} F_n \left( A'_{n,m} r^{np} + C'_{n,m} r^{-np} - \frac{\delta_{m,2}\, r^2}{4-\left(np\right)^2}\right) \cos (np \theta).
\end{align}

\section{Coefficients for Open-Circuit Field}
\label{coefficients-open-circuit}
\begin{align}
A_{n,1} = &K^{-1} R_{im}^{-np} \left(R_{im}^{2np+1}\left(R_y^{2np}+R_{om}^{2np}\right) (1-np)(1-\mu_{r,2}) + R_{im} R_{om}^{4np}(1+np)(1-\mu_{r,2}) \right. \nonumber \\
&\left.+ R_{im} R_y^{2np} R_{om}^{2np} (1+np)(1+\mu_{r,2}) + 2 R_{im}^{np} R_{om}^{3np+1} (np\mu_{r,2} - 1) - R_{om}^{np+1} R_y^{2np} R_{im}^{np} (np \mu_{r,2} +1) \right),\\
C_{n,1} = & 0, \\
A_{n,2} = & (\mu_{r,2} K)^{-1} \left(R_y^{2np} R_{im}^{np+1}(1 - \mu_{r,2}) (1 - np\mu_{r,2}) + R_{om}^{3np+1} (-1-\mu_{r,2})(1-np\mu_{r,2}) \right. \nonumber \\
&\left.- R_y^{2np}R_{om}^{np+1} (1 + \mu_{r,2}) (1 + np\mu_{r,2}) + R_{im}^{np+1} R_{om}^{2np} (1 + \mu_{r,2}) (1 - np\mu_{r,2})\right), \\
C_{n,2} = & (\mu_{r,2} K)^{-1}R_{im}^{np}R_{om}^{np} \left(R_y^{2np}R_{im} R_{om}^{np} (-1 -\mu_{r,2})(1-np\mu_{r,2}) + R_{im}^{np}R_{om}^{2np+1} (1-\mu_{r,2})(1-np\mu_{r,2})  \right. \nonumber \\
&\left. + R_y^{2np} R_{im}^{np} R_{om} (1- \mu_{r,2})(1+np\mu_{r,2}) + R_{im} R_{om}^{3np}(-1+\mu_{r,2})(1-np\mu_{r,2})\right), \\
A_{n,3} = & K^{-1} \left( R_{im}^{2np} R_{om} (-1 + \mu_{r,2})(1+ np) + 2 R_{im}^{np+1} R_{om}^{np} (1-np \mu_{r,2}) + R_{om}^{2np+1} (-1 -\mu_{r,2})(1-np) \right), \\
C_{n,3} = & K^{-1} R_y^{2np} R_{om}^{np} \left( 2 R_{im}^{np+1} R_{om}^{np} (1+np\mu_{r,2}) + R_{im}^{2np} R_{om} (1-\mu_{r,2})(1+np) + R_{om}^{2np+1} (1+\mu_{r,2})(1-np) \right), \\
A_{n,4} = &0, \\
C_{n,4} = &0, \\
K = &(np+1)(np-1) \left( R_y^{2np}R_{im}^{2np} (1-2\mu_{r,2}+\mu_{r,2}^2) \right. \nonumber \\ & \left. - R_y^{2np}R_{om}^{2np} (1+2\mu_{r,2}+\mu_{r,2}^2) + (R_{im}^{2np}R_{om}^{2np} - R_{om}^{4np})(1-\mu_{r,2}^2) \right).
\end{align}

\section{Coefficients for Armature Reaction Field}
\label{coefficients-armature-reaction}
\begin{align}
A'_{n,1} = &E^{-1} \mu_{r,2} \left(R_y ^{2np}R_{iw}^{2np+2} (-1+\mu_{r,2})(2-np) + R_{ow}^{np+2} R_y^{2np}R_{iw}^{np} (np+2) + R_{iw}^2R_{ow}^{4np}(-1+ \mu_{r,2})(2+np)  \right. \nonumber \\
&\left. - R_{iw}^2 R_y^{2np} R_{ow}^{2np}(1+\mu_{r,2})(2+np) + 2 R_{iw}^{np} R_{ow}^{3np+2} (2-np) + R_{iw}^{2np+2} R_{ow}^{2np} (-1-\mu_{r,2})(2-np) \right), \\
C'_{n,1} = &0, \\
A'_{n,2} = &E^{-1}\left( R_y^{2np}R_{iw}^{np+2}(-1+\mu_{r,2})(2-np\mu_{r,2}) + R_{ow}^{3np+2}(1+\mu_{r,2})(2-np\mu_{r,2}) \right. \nonumber \\
& \left. + R_y^{2np}R_{ow}^{np+2}(1+\mu_{r,2})(1+np\mu_{r,2}) + R_{iw}^{np+2}R_{ow}^{2np}(1+\mu_{r,2})(-2+np\mu_{r,2}) \right), \\
C'_{n,2} = &E^{-1} R_{iw}^{np}R_{ow}^{np} \left( R_{iw}^2R_{ow}^{3np}(-1+\mu_{r,2})(2-np\mu_{r,2}) + R_y^{2np} R_{iw}^{np} R_{ow}^2 (1- \mu_{r,2})(2-np\mu_{r,2}) \right. \nonumber \\
&+ R_{iw}^2 R_y^{2np}R_{ow}^{np} (-1-\mu_{r,2})(2-np\mu_{r,2}) + R_{iw}^{np}R_{ow}^{2np+2}(1-\mu_{r,2})(2-np\mu_{r,2}), \\
A'_{n,3} = & E^{-1} \mu_{r,2} R_{ow}^{np} \left( 2 R_{iw}^{np+2} R_{ow}^{np} (-2+np) + R_{iw}^{2np} R_{ow}^2 ( 1- \mu_{r,2})(2+np) + R_{ow}^{2np+2} ( 1+\mu_{r,2})(2-np)\right), \\
C'_{n,3} = & E^{-1}R_y^{2np}R_{ow}^{np} \left( R_{iw}^{np+2}R_{ow}^{np} (-2+np\mu_{r,2}) + R_{iw}^{2np} R_{ow}^2 ( 1-\mu_{r,2})(2+np) + R_{ow}^{2np+2}(1-\mu_{r,2})(2+np)\right), \\
A'_{n,4} = &0, \\
C'_{n,4} = &0, \\
E = &np(np+2)(np-2) \left( R_y^{2np}R_{iw}^{2np} (1-2\mu_{r,2}+\mu_{r,2}^2) \right. \nonumber \\ & \left. - R_y^{2np}R_{ow}^{2np} (1+2\mu_{r,2}+\mu_{r,2}^2) + (R_{iw}^{2np}R_{ow}^{2np} - R_{ow}^{4np})(1-\mu_{r,2}^2) \right).
\end{align}

\end{document}